\documentclass[pre,amsmath,amssymb,superscriptaddress,showpacs]{revtex4-1}

\usepackage{amsmath,amssymb,graphicx}
\usepackage{enumitem}
\usepackage{amsbsy}
\usepackage{latexsym}
\usepackage{color}
\usepackage{graphicx}
\usepackage{psfrag}
\usepackage[normalem]{ulem}
\usepackage{bm}

\newcommand{\be}{\begin{equation}}
\newcommand{\ee}{\end{equation}}
\newcommand{\bea}{\begin{eqnarray}}
\newcommand{\eea}{\end{eqnarray}}

\newcommand{\comment}[1]{}

\begin{document}

\title{Thermalization with a multibath: an investigation in simple models}

\author{Giovanni Battista Carollo}
\email{giovanni.carollo@uniba.it}
\affiliation{Dipartimento di Fisica, Universit\`a degli Studi di Bari and INFN,
Sezione di Bari, via Amendola 173, Bari, I-70126, Italy}
\author{Federico Corberi}
\email{corberi@sa.infn.it}
\affiliation{Dipartimento di Fisica ``E.~R. Caianiello'' and INFN, Gruppo Collegato di Salerno,
and CNISM, Universit\`a di Salerno, via Giovanni Paolo II 132, 84084 Fisciano (SA), Italy.}
\author{Giuseppe Gonnella}
\email{giuseppe.gonnella@ba.infn.it}
\affiliation{Dipartimento di Fisica, Universit\`a degli Studi di Bari and INFN,
Sezione di Bari, via Amendola 173, Bari, I-70126, Italy}

\begin{abstract}
We study analytically and numerically a couple of paradigmatic spin models, each described in terms of two sets of variables attached to two different thermal baths with characteristic timescales $T$ and $\tau$ and inverse temperatures $B$ and $\beta$.
In the limit in which one bath becomes extremely slow ($\tau \to \infty$), such models amount to a paramagnet and to a one-dimensional ferromagnet, in contact with a single bath. We show that these systems reach a stationary state in a finite time for any choice of $B$ and $\beta$. We determine the non-equilibrium fluctuation-dissipation relation between the autocorrelation and the response function in such state and, from that, we discuss if and how thermalization with the two baths occurs and the emergence of a non-trivial fluctuation-dissipation ratio.
  
\end{abstract}

\maketitle

\section{Introduction}
\label{intro}

Equilibrium statistical mechanics deals with systems which are either isolated or exchange  
heat, particles, or other quantities with a single external reservoir or bath. Well known cases are the canonical and the grand-canonical ensembles, where either heat (alone) or heat and particles are exchanged and the amount of them retained by
the system is tuned by bath intensive parameters, temperature and chemical potential. As long as
a single bath is present, an equilibrium state exists which, if ensemble equivalence holds, is unique
and independent of the presence of the reservoir.

When different baths are present the system is found, generally, in a 
non-equilibrium state. This case will be denoted in the following 
as a {\it multibath}, in contrast to the usual single bath setting. There are several physical situations of 
this kind, very widespread in nature and of 
great practical importance, such as 
systems between two temperatures \cite{reviewRayleighBenard,Lepri_2003,Derrida_2005_proc,Derrida_2005,Lepri_1896,Cugliandolo_1999,Cugliandolo_1999_1,Jarzynski_2004,Lecomte_2005,Visco_2006, Piscitelli_2008, Piscitelli_2009,Crisanti_2012, Borchers_2014,Crisanti_2012} or in contact with two particle reservoirs   
\cite{Schmittmann_1998,traffic1,traffic2,Popkov}. Indeed, the latter problem
has been widely studied theoretically as paradigmatic for non equilibrium systems \cite{MacDonald1968,Spitzer,Evans_2000,Evans_2005,Bodineau_2006,Brzank,Cohen_2008, Zia_2011,Essler}. In the
following we will always consider the case where the multibath 
exchanges heat (alone) with the sample. 

Even with a single reservoir the existence of a stationary state is not 
guaranteed because there are systems which do not attain equilibration in finite times in the thermodynamic limit. If brought away from equilibrium they age forever, examples of which are several instances of magnets and of glassy systems~\cite{struik1978,Lundgren,Cugliandolo_1994,Bouchaud1,Bouchaud2}. This sometimes introduces serious limitations on the experimental and numerical study of the equilibrium state, which can only be obtained for small system sizes. In this respect, putting such aging systems in contact with a multibath may unexpectedly represent an advantage, since in many cases this forces the system to stationarize. Physically, this occurs because the low-energy, slow modes informing the aging states are destabilized by the fluxes induced by the multibath inside the system.
Let us mention, by the way, that there are also cases where this has been shown not to occur~\cite{Oliveira_1993,Andrenacci_2006}.
Then, if a vocabulary exists to translate the stationary properties of multibath models to the equilibrium ones of single bath systems, one can more easily study the former to understand the latter. 

As we shall discuss now, such vocabulary exists in some special cases because non-equilibrium stationary states of multibath models may, in some conditions, correspond to equilibrium states of dual disordered systems {\cite{Contucci_2019_1,Contucci_2021,Alberici_2022}. Take for instance a spin system, with a finite equilibration time, subjected to external quenched random magnetic fields and in contact with a single bath. This system can be transformed in a dual multibath one by promoting the quenched magnetic fields to dynamic degrees of freedom evolving on a timescale $\tau$ in contact with a second bath at inverse temperature $\beta$.  In the limit in which $\tau$ is very large and $\beta $ is very small, its effect will be, once the spins have equilibrated with it, to randomly reshuffle the values of the magnetic fields. Hence, the thermodynamic of such multibath system will amount to that of the model with a single bath and the reshuffling of the random field realizes the average over the quenched randomness (see Sec.~\ref{models} for a more precise statement). This is expected to be true quite in general and discloses the possibility of studying the statics of the equilibrium states of disordered systems, where a quenched average over the disorder has to be performed, in terms of the stationary state they attain when put in contact with a multibath. In addition to that, other properties of such stationary states are interesting, since it has been argued~\cite{Contucci_2021} that some of the investigations that one usually does with a single bath, such as thermodynamic integration and the analysis of the dynamic fluctuation-dissipation relation, can be extended to the multibath case.

Despite the potential interest of the issues discussed insofar, a clear and well-established set of affordable results on these subjects is, in our opinion, still lacking. Therefore, in this paper we tackle some of these questions in simple statistical-mechanical models in contact with a multibath where analytical calculations are fully or partially doable and numerical simulations provide clear-cut evidence. 
Specifically, we consider two paradigmatic models whose dual disordered counterparts amount to a 
non-interacting paramagnet and to an interacting spin-glass, later described also as model i) and ii).   

In this framework we study the issue of the stationarization and consider the way the models thermalize with the multibath by inspection of the fluctuation-dissipation relation, discussing also the meaning of the effective temperatures that can be extracted from such relation~\cite{Cugliandolo_2011, Cugliandolo_1994, Cugliandolo_1997,Corberi_2005_shear,Petrelli_2020}.
This will allow us to show how the presence of the interactions in model ii) may enrich and 
complicate the simple and intuitive physics found in model i), leading to a fully non-trivial fluctuation-dissipation ratio, akin to the one found in one-dimensional aging ferromagnets on small timescales and taking a constant value on larger timescales. 

This paper is organized as follows:
in Sec.~\ref{models} we set the notation, define the models we will study, discuss the invariant measure of a multibath system and its thermalization properties.
In Sec.~\ref{paramagnet} we study analytically how these properties are realized in the simple, solvable paramagnetic model i). In Sec.~\ref{interacting} we carry out a similar analysis in the interacting system ii). Finally, in Sec.~\ref{conclusions} we recapitulate what we found, discuss the results and point out some open problems.

\section{Models, timescales and thermalization} \label{models}

We consider systems described by 
two sets of $N$ Boolean variables, $\{S_i \}$
($S_i=\pm 1, \forall i$) and 
$\{\sigma_i\}$ ($\sigma_i=\pm1, \forall i$), in contact with 
two baths at inverse temperatures 
$B$ and $\beta$, which act on
different timescales $T$ and $\tau$. 
Notice that, in order to make the notation transparent, quantities (variables and
parameters) associated to the fast evolution (like $S_i$, $B$, $T$) are in latin letters, while those 
associated to the slow evolution are in greek letters. Systems of this kind, where both type of variables are evolving, will be denoted as {\it annealed}. We will always consider the case $B\ge \beta$ and $T\ll \tau$.

Let us also define the associated {\it quenched} versions, represented by the same systems where, however,
the $\{\sigma_i\}$ are random variables which do not vary in time and are extracted from a prior distribution
$p(\{\sigma_i\})$ which will be assumed to be flat, i.e. $p(\{\sigma_i\})=2^{-N}$.

Coming back to the annealed systems (these are the ones we will always consider in the following, unless explicitly mentioned),
a true equilibrium exists for $B=\beta$ and, in this case, the free energy reads
\be
\beta F=-\ln \left (\sum _{\{S_i\},\{\sigma _i\}}e^{-\beta {\cal H}(\{S\},\{\sigma_i\})}\right ),
\label{ann}
\ee
where ${\cal H}(\{S_i\},\{\sigma _i\})$ is the Hamiltonian, which holds true for any $T,\tau$. Stationary states with $B\neq \beta$ cannot be equilibria, because some net heat will flow between the $S_i$ and the $\sigma_i$, breaking time-reversal symmetry. However, for $\tau\gg T$ the stationary state 
can still be studied, possibly, with the methods of equilibrium statistical mechanics because such states are expected to correspond to the equilibrium ones of the associated quenched systems, the variables $\{\sigma _i\}$ not changing in the
time needed to the $\{S_i\}$ to relax.
Indeed, exploiting this, we first write the {\it free energy} at fixed $\{\sigma_i\}$ as
\be
F_\sigma(\{\sigma_i\})=-B^{-1}\ln\left (\sum _{\{S_i\}}e^{-B {\cal H}(\{ S_i\},\{\sigma _i\})}\right).
\ee 
The quantity $e^{-\beta  F_\sigma(\{\sigma_i\})}$ describes the (not normalised) probability of having a particular realisation of the $\{\sigma_i\}$, so that we can write
the free energy of the annealed system as
\be
F=-\beta^{-1}\ln \sum_{\{\sigma_i\}}e^{-\beta  F_\sigma(\{\sigma_i\})}=-\beta^{-1}\ln\left [\sum_{\{\sigma_i\}} \left (\sum _{\{S_i\}}e^{-B {\cal H}(\{S_i\},\{\sigma_i\}}\right )^n\right ]=-B^{-1}\frac{\ln\overline {Z_\sigma^n}}{n}
,
\label{nested}
\ee
where $Z_\sigma=\sum _{\{S_i\}}e^{-B {\cal H}(\{S_i\},\{\sigma_i\})}$, $n=\frac{\beta}{B}$ and, in the last passage, we have recognised the quenched average over the $\{\sigma_i\}$ randomness $\overline {(\cdots)} =\sum_{\{\sigma_i\}} p(\{\sigma_i\})(\cdots)$ (the constant factor $p(\{\sigma_i\})$ has
been omitted in Eq.~(\ref{nested})). 
We will call this the {\it nested} structure of $F$, since fast variables are traced over before the slow ones~\cite{Contucci_2019_1,Contucci_2021}.
For $n  =1$ one recovers the standard equilibrium expression~(\ref{ann}).
Conversely, for $n \to 0$,   
in the last term of above equation one recognizes the replica receipt~\cite{mezard1987spin}
to compute the free energy of the associated quenched system.

In this paper
we will discuss the two models defined by the Hamiltonians:
\begin{enumerate}[label=\roman*)]
    \item 
 \begin{equation}
        {\cal H}(\{S_i\},\{\sigma_i\})=-\sum_i
        S_i\sigma_i.
        \label{ham1}
    \end{equation}
    Recalling the previous discussion, the associated quenched 
    model is a paramagnet in a magnetic field $\sigma_i$. For this reason in the following we will also use the word {\it paramagnet} (or {\it non-interacting}) to refer to this system. Let us stress, however,
    that the annealed model is neither a paramagnet neither non-interacting. The same terminological caution will apply to the following model

    \item   
 \begin{equation}
        {\cal H}(\{S_i\},\{\sigma_i\})=-\sum_i
        \sigma_i S_iS_{i+1}.
        \label{ham2}
    \end{equation}
    The associated quenched model is a one-dimensional Ising system, the $\{\sigma_i\}$ playing the role of random coupling constants.  It is well known that a one-dimensional spin-glass can be mapped on a simpler ferromagnetic system~\cite{Nishimori}. However, in order to do that, the $\sigma_i$ must be independent variables which, in the present case, is only true for $\beta=0$. We will denote this system as a spin-glass (or {\it interacting} model).
\end{enumerate}

The above models are complemented by a stochastic kinetics where single variables $S_i$ or $\sigma _i$ flip with transition rates 
\be
w_{S_i}(\{ S_i\},\{\sigma _i\})=\frac{1}{NT}\,\omega_{S_i}({\{S_i\},\{\sigma _i\}})\quad,\quad
w_{\sigma_i}(\{S\},\{\sigma_i\})= \frac{1}{N\tau}\,\omega _{\sigma_i}(\{ S_i\},\{\sigma_i\}). 
\ee  
$\omega _{S_i}$ and $\omega_{\sigma_i}$  obey detailed balance with respect to the Hamiltonians~(\ref{ham1}-\ref{ham2}), 
with inverse temperatures $B$ and $\beta$ respectively. They will be chosen of the Glauber type~\cite{Glauber}
\be
\omega _{S_i}(\{ S_i\},\{\sigma_i\})=
\frac{1}{2}\text {sech}\bigg[\frac{B}{2}\Delta E(\{S_i\},\{\sigma_i\})\bigg]
e^{-\frac{B}{2}\Delta E(\{S_i\},\{\sigma_i\})},
\label{rate1}
\ee
\be
\omega_{\sigma_i}(\{ S_i\},\{\sigma_i\})=
\frac{1}{2}\text {sech} \bigg[\frac{\beta}{2}\Delta E(\{S_i\},\{\sigma_i\})\bigg]
e^{-\frac{\beta}{2}\Delta E(\{S_i\},\{\sigma_i\})},
\label{rate2}
\ee
where  $\Delta E$ is the energy variation due to a flip.

As it will be shown, the two models considered attain a stationary state after a finite time $t_{stat}$ (if prepared far from it). This is expected for a model as simple as i), but is not obvious for system ii). Indeed, its quenched version is an interacting model that ages forever at $B=\infty$ in the thermodynamic limit, hence $t_{stat}=\infty$ (for
large but finite $B$ there is an interrupted, but long-lasting, aging~\cite{Lippiello-Zannetti, Corberi_2019, Corberi_2019b}). 

Let us point out that, despite the close relation discussed above between the static properties 
of the annealed model with $\tau\gg T$ (where the nested structure~(\ref{nested}) is expected to hold) and the quenched one, the two systems may differ significantly in their dynamical aspects. As we will see, this is true even
in the stationary states considered
in this paper.
Such states, by definition, are observed on times $t/t_{stat}=$ const$>1$. In the relevant regime $\tau\gg T$, if $t_{stat}\gtrsim \tau$ -- which holds true for model ii) -- the system may take a huge time to attain stationarity.

The properties of the stationary states are informed by
the presence of different timescales,
which we briefly discuss now in the case $\tau\gg T$.
In this limit the relaxation time $t_S$ of the variables $\{S_i\}$
in the quenched model
can be inferred from the decay of the autocorrelation function 
\begin{equation}
    C(t,s)=C(s)=\langle S_i(t)S_i(t+s)\rangle
    \label{defC}
\end{equation}
computed in the annealed system, restricting to times $s\ll \tau$ .
For instance, in model i) the relaxation time of the spins $\{S_i\}$ in the quenched model is, in general, of order $T$ and in this time domain $C(s)$ decays from $C(s=0)=1$ to 
$C(s\simeq t_S)\simeq m^2$, where $m^2=(\tanh B)^2$ is the squared equilibrium magnetization. 
By the way, notice that $t_S$ may be ill defined if the equilibrium state is frozen, which is 
the case at $B=\infty$, because
there is no decay of $C(s)$ in this time domain. This fact will have  consequences when discussing the fluctuation-dissipation relation and the thermalization properties. 

From the decay of $C(s)$ on much larger timescales, namely for $s\gtrsim \tau\gg t_S$,
one can infer a further relaxation time $t_{\sigma S}$, which is associated to the
rearrangements of the $\{S_i\}$ as due to the slow evolution of the $\{\sigma _i\}$. For instance, in model i), $C(s)$ in this time domain decays from $C(s\simeq t_S)=m^2$ to $C(s\simeq t_{\sigma S})\simeq 0$, because the local magnetizations $\langle S_i\rangle$ are reshuffled by the modifications of the {\it magnetic fields} $\sigma _i$. Clearly, in the case of a model as simple as i) it is $t_{\sigma S}\propto \tau$.

The timescales discussed above are fundamental to understand the thermalization properties. This can be done in terms of the fluctuation-dissipation relation (FDR).
In an equilibrium system in contact with a unique thermal bath
at inverse temperature ${\cal B}$, the fluctuation-dissipation theorem~\cite{Kubo_1966}
\begin{equation}
    R(s)=-{\cal B}\,\frac{dC(s)}{ds}
    \label{fdt}
\end{equation}
relates the autocorrelation function with the response function 
\begin{equation}
    R(s)=\left .\frac{\partial \langle S_i \rangle}{\partial h_i}\right \vert_{h_i=0},
    \label{defR}
\end{equation}
where $h_i$ is a field linearly coupled with $S_i$ in the Hamiltonian, i.e., in the models we deal with, a (quenched) magnetic field. In Eq.~(\ref{fdt}), the bath temperature enters as a proportionality constant.
In the presence of a multibath with well separated timescales (i.e. $\tau\gg T$ in our bi-bath systems) we will say that the system is multi-thermalized with the baths if, on each bath characteristic timescale, one has a relation as Eq.~(\ref{fdt}) with ${\cal B}$ replaced by the inverse temperature of the reservoir associated to that particular timescale~\cite{Cugliandolo_2011}. For instance, in our models, if multi-thermalization occurs, Eq.~(\ref{fdt}) should be obeyed with ${\cal B}=B$ in the timesector $s\lesssim t_S$ and with ${\cal B}=\beta$ for $s\gtrsim t_{\sigma S}$.
We will see that this actually happens, except in some interesting special cases. Let us remark that, in our acceptation, thermalization, or multi-thermalization, are always meant with respect to the baths, we will not deal with the more complicated subject of internal thermalization among different degrees of freedom (for instance between and the $\{S_i\}$ and the $\{\sigma _i\}$).
Also, in this paper we will be interested only in the thermalization properties of the {\it fast} variables $\{S_i\}$, as it is obvious since $C$ and $\chi$ are correlations and response functions of such quantities.

\section{Model i): Paramagnet} \label{paramagnet}

From now on we will set, without loss of generality, $T=1$ (this way, $\tau$ represents the relative speed of the slow bath with respect to the fast one). The average energy in this case is given by $E(t)=-\langle \sigma_i(t) S_i(t) \rangle$ and, through a standard calculation (see appendix \ref{apppara}), it evolves according to
\be
\frac{dE(t)}{dt}=-\left(1+\frac{1}\tau\right)E(t)+m+\frac{\mu}\tau \ ,
\label{evoenergy}
\ee
where $m=\tanh B$ and  $\mu=\tanh \beta$. Assuming all the Boolean variables initialized randomly (with null average), the solution is
\be
E(t)= -\frac{m+\frac{\mu}\tau}{1+\frac{1}\tau}\left(1-e^{-t\left(1+\frac{1}\tau\right)}\right),
\ee
showing that the system stationarizes on a time $t_{stat}=\frac{\tau}{\tau+1}$.
This implies that there exists a characteristic timescale $\tau _c=1$
such that, for $\tau\ll \tau_c$ it is
$t_{stat}\simeq \tau$, whereas $t_{stat}\simeq 1$ for $\tau \gg\tau_c$.
Notice that $\tau_c=T=1$ is the stationarization time of the associated quenched model. This means that, as long as $\tau <\tau_c$, stationarization is induced by the reshuffling of the $\{\sigma _i\}$,
whilst for $\tau >\tau_c$ it is due 
to the flipping of the $\{S_i\}$. Since in this paper we are interested in the case $\tau > T=1$, the existence of two stationarization mechanisms as $\tau$ is varied is not
particularly relevant in this model. However, we will see that the same phenomenon occurs in model ii) where, in contrast, $\tau_c$ can be much larger than $T$.

An analogous computation (see appendix \ref{apppara}) leads to the following equations for the autocorrelation function defined in Eq.~(\ref{defC})
\begin{eqnarray}
    \frac{\partial C(t,s)}{\partial s}&=&-C(t,s)+mA(t,s) \nonumber \\
    \tau\frac{\partial A(t,s)}{\partial s}&=&-A(t,s)+\mu C(t,s),
    \label{sistC}
\end{eqnarray}
where $A(t,s)=\langle S_i(t)\sigma_i (t+s)\rangle$.
Solving these equations with the appropriate initial conditions (that are $C(t,0)=1$ and $A(t,0)=-E(t)$) one finds
(see Appendix~\ref{apppara}) that $C(t,s)$ becomes a function of $s$ alone after a time of order $t \simeq 1$.
In this time domain the expression for $C$ reads
\begin{equation}
 C(s)=(1-m^2)e^{-s/t_S}+m^2e^{-s/t_{\sigma S}}, 
 \label{Cstat}
\end{equation}
where 
$
t_S= \frac{2\tau}{1+\tau+\sqrt{(\tau-1)^2+4\tau m \mu}}\simeq 1$ and
$t_{\sigma S}= \frac{2\tau}{1+\tau-\sqrt{(\tau-1)^2+4\tau m \mu}}\simeq \frac{\tau}{1-m \mu}$, the last expressions (after $\simeq$) holding for large $\tau$.
This shows that the relaxation times $t_S$ and $t_{\sigma S}$
are comparable to the timescales of the two baths, as expected. Since $m$ is the equilibrium magnetization of the model with the quenched 
$\{\sigma _i\}$, Eq.~(\ref{Cstat}) transparently shows that $C$ decorrelates down to $\langle S_i(t)\rangle \langle S_i(t+s)\rangle=m^2$ on the fast timescale $t_S$ and then, on the much longer timescale $t_{\sigma S}$, it happens that $\langle S_i(t+s)\rangle\to 0$ due to the further decorrelation caused by the evolution of the $\{\sigma _i\}$ and hence $C(s)$ decays to zero. 

Next we consider the response function. Using the generalization of the fluctuation-dissipation theorem to non-equilibrium states derived in~\cite{Lippiello_2005,Corberi_2007,PhysRevB.77.212201,PhysRevE.78.041120,Baiesi_2009,PhysRevE.81.011124}, in the stationary state we have
\be
B^{-1}R(s)=-\frac{dC(s)}{ds}-
\frac{m}{2}\left [\widetilde A(s)-A(s)\right ],
\label{resp}
\ee
where $\widetilde A(s)=\langle \sigma _i(t)S_i(t+s)\rangle$ which,
proceeding as to arrive at Eqs.~(\ref{sistC}), obeys
\begin{eqnarray}
 \frac{\partial \widetilde A(t,s)}{\partial s}&=&-\widetilde A(t,s)+m \Sigma(t,s) \nonumber \\
  \tau\,\frac{\partial  \Sigma (t,s)}{\partial s}&=&- \Sigma(t,s)+\mu\widetilde A(t,s),
  \label{sistA}
  \end{eqnarray}
where $\Sigma (t,s)=\langle \sigma_i(t)\sigma _i(t+s)\rangle$.
These equations are identical to Eqs.~(\ref{sistC}), with obvious substitutions, so we have for $\widetilde A$ the same expression~(\ref{Cstat}) previously discussed for $C$, at stationarity. 

All the functions appearing on the r.h.s. of Eq.~(\ref{resp})
can be computed analytically solving the system~(\ref{sistC})
(and the equivalent~(\ref{sistA})).
Hence one can explicitly compute
\begin{equation}
    X(s)=-\frac{R(s)}{B\frac{dC(s)}{ds}},
\end{equation}
finding
\be
X(s)=\frac{a_1\,e^{-s/t_S}+a_2\,e^{-s/t_{\sigma S}}}{a_3\,e^{-s/t_S}+a_4\,e^{-s/t_{\sigma S}}}
\ee
with
$a_1=\left [\frac{1-m^2}{t_S}+\frac{m^2\left(1-\frac{\beta}{B}\right)}{t_{\sigma S}}\right ]$,
$a_2=\frac{\beta}{B}\frac{m^2}{t_{\sigma S}}$,
$a_3=\frac{1-m^2}{t_S}$ and
$a_4=\frac{m^2}{t_{\sigma S}}$.
For large $\tau$, when $t_S\ll t_{\sigma S}$, one has
\be
X(s)=\left \{
\begin{array}{lll}
1\quad , &\mbox{for} &s\ll t_{\sigma S} \\
\frac{\beta}{B}\quad , &\mbox{for} &s\gtrsim t_{\sigma S}.
\end{array}
\right .
\label{Xs}
\ee
Parametrizing $s$ in terms of $C$, using Eq.~(\ref{Cstat}), in the limit $\tau \rightarrow \infty$ one obtains also
\be
X(C)=\left \{
\begin{array}{lll}
\frac{\beta}{B}\quad , &\mbox{for} &0\le C\le m^2 \\
1\quad , &\mbox{for} &m^2\le C\le 1.
\end{array}
\right .
\label{lateX}
\ee
Recalling the discussion on multi-thermalization put forward in Sec.~\ref{models}
(below Eq.~(\ref{defR})),
this equation (or, similarly, Eq.~(\ref{Xs})), shows that the system is multi-thermalized with the two baths at inverse temperatures $B$ and $\beta$ on the two, well separated, characteristic timescales of the reservoirs.

Instead of considering $R$, particularly in simulations or experiments, it is usual to study the {\it integrated} response function, or dynamical susceptibility
\begin{equation}
    \chi(s)=\int _0^sR(s')\,ds',
    \label{chi}
\end{equation}
which, in an equilibrium state at inverse temperature ${\cal B}$, obeys
\be
{\cal B}^{-1}\chi(s)=1-C(s).
\label{equiR}
\ee
This quantity is plotted against $C$ in Fig.~\ref{fig_fdtplot_paramagnet}.
In this representation $X(C)$ is 
the slope of the curve.
In the left panel, besides showing the equilibrium line~(\ref{equiR}) as a guide to the eye, we plot the curves for fixed values of $\beta=0.5$ and $B=1$, for different values of $\tau$. We see that the analytical form~(\ref{lateX}) 
is recovered for large $\tau$ (e.g., for $\tau=10^2$). The effect of a finite $\tau$ is to round the curve, the more the smaller $\tau$ is, producing a spurious slope.
On the right panel we consider the role of changing $\beta$, for fixed $B=1$ and
for a large value of $\tau=10^2$. Also in this conditions, the limiting form~(\ref{lateX}) is well reproduced, for any $\beta$.

\begin{figure}[htbp]
	\centering
        \includegraphics[width=0.49\textwidth]{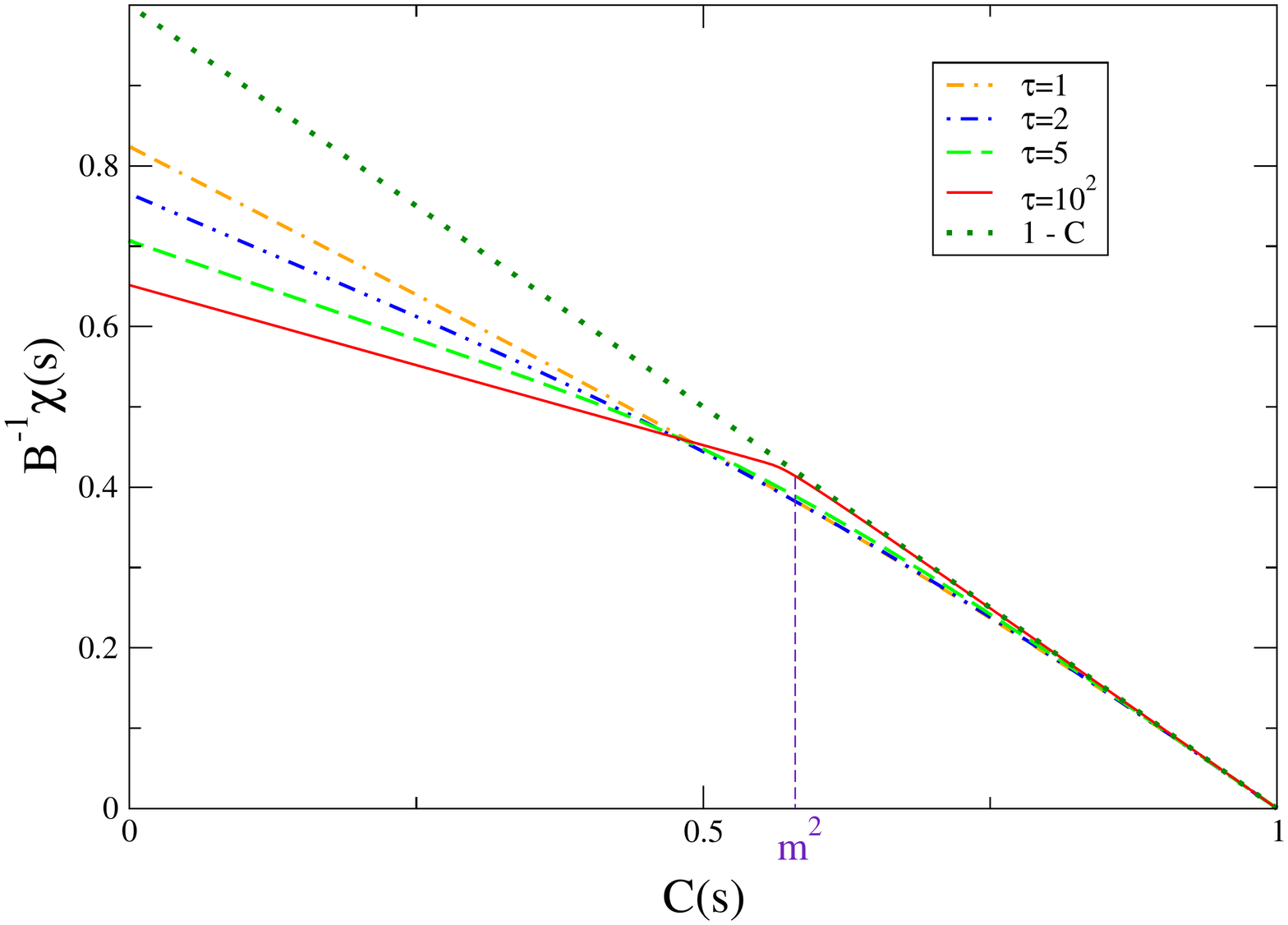}
        \includegraphics[width=0.49\textwidth]{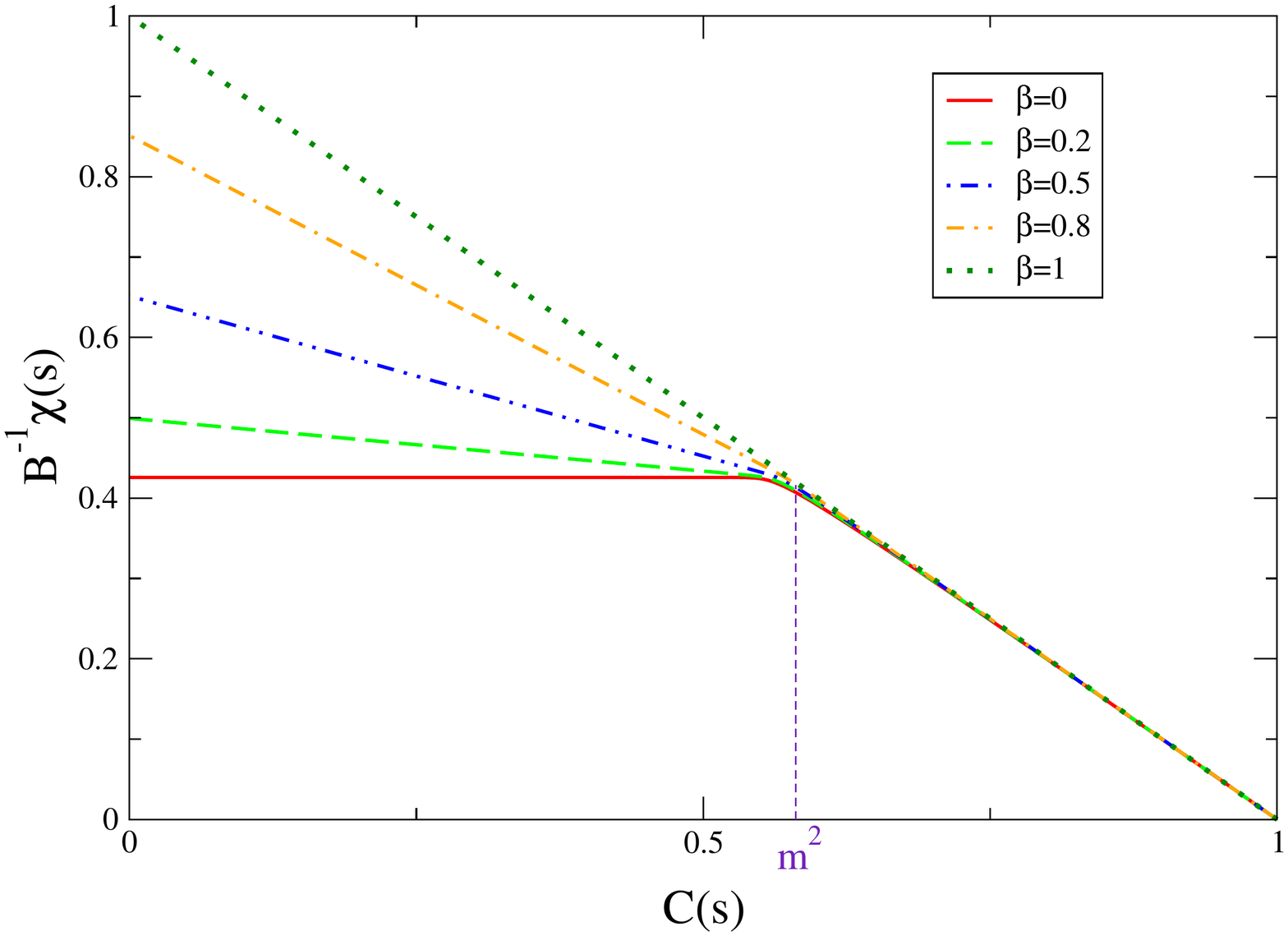}	
	\caption{$B^{-1}\chi(s)$ is plotted against $C(s)$ for $\beta=0.5$ and different values of $\tau$ (left panel) or for $\tau=10^2$ and different values of $\beta$ (right panel). In both cases $B=1$. 
	The dotted green line is the equilibrium form~(\ref{equiR}). The vertical line corresponds to $C(s)=m^2$. The curves have been obtained by computing the integrated response analytically, as explained in Appendix \ref{apppara}.}
	\label{fig_fdtplot_paramagnet}
\end{figure}

Finally, let us now consider the case $B=\infty$, which needs a separate discussion. Now the quenched model is frozen, 
hence there is no decay of $C(s)$, meaning that $C(s)\equiv 1,\, \forall s$.
In the annealed model this is reflected by the fact that the first term of Eq.~(\ref{Cstat}) vanishes, since $m^2=1$ as $B=\infty$, giving
\begin{equation}
C(s)=e^{-\frac{s}{t_{\sigma S}}}
\end{equation}
in this case.
Similarly, in the limit $B\to \infty$, one easily obtains
\begin{equation}
    X(s)=e^{-s\left (t_S^{-1}-t_{\sigma S}^{-1}\right )}\simeq e^{-\frac{s}{t_S}}\simeq
    \left \{
\begin{array}{lll}
1\quad , &\mbox{for} &s\ll t_ S \\
0\quad , &\mbox{for} &s\gg t_S,
\end{array}
\right .
\end{equation}
the second passage holding for large $\tau$. This shows once again multi-thermalization of the model with the two baths.
Re-parametrizing in terms of $C$
one finds
\begin{equation}
    X(C)=C^{\frac{t_{\sigma S}}{t_S}}=\left \{
    \begin{array}{lll}
    0\quad , &\mbox{for} &0\le C<1 \\
1\quad , &\mbox{for} & C= 1,
\end{array}
\right .
\label{X_para_T0}
\end{equation}
the last passage holding for $\tau\to \infty$.
Multi-thermalization is clearly still there but, in this representation, the role of the fast bath can be hardly recognized since it is reduced to a single point ($C=1$) in the diagram of $\chi$ vs $C$.

\section{Model ii): Spin-glass} \label{interacting}

In this section we study model ii) which, at variance with the previous one, is not an analytically exactly solvable. However, it is still possible to develop some approximation scheme or numerical analysis.

Let us start with the issue of stationarization, showing that it holds for any value of $B$, including $B=\infty$. As we mentioned already, this result is not trivial because the corresponding quenched model stationarizes only if $B$ is finite. 
This property can be inferred from the behavior of the average energy
\begin{equation}
    E(t)=-\langle \sigma _iS_iS_{i+1}\rangle ,
    \label{energy}
\end{equation}
which, proceeding again as in~\cite{Glauber} (see Appendix \ref{appInt}), obeys
\begin{equation}
\frac{dE(t)}{dt} =
-2\left (1+\frac{1}{2\tau}\right )E(t) 
-u[1+G_2(t)]-\frac{\mu}{\tau},
\label{eqE}
\end{equation}
 where $G_2(t)=\langle S_{i}\sigma_{i}\sigma_{i+1}S_{i+2}\rangle$ is a four-variables correlation and $u=\tanh(2B)$.  
 In order to obtain a closed equation one can, for instance, devise the following approximation scheme.
 We insert $S_i^2=1$ in the correlator defining $G_2(t)$ and split it as follows
 \be
 G_2=\langle S_{i-1}\sigma_{i-1}\sigma_iS_{i+1}\rangle=\langle S_{i-1}\sigma_{i-1}S_iS_i\sigma_iS_{i+1}\rangle \quad \longrightarrow\quad \langle S_{i-1}\sigma_{i-1}S_i\rangle \langle S_i\sigma_iS_{i+1}\rangle =E^2.
 \label{split}
 \ee
It can be shown (see Appendix \ref{appInt}) that the approximation becomes exact for $\tau \to \infty$ or for  $\beta \rightarrow \infty$, $B \rightarrow \infty$, which are the kind of limits we are mostly interested in. Furthermore, the approximation scheme can be improved upon closing the equations at the level of $N$-variables
correlators, instead of working with a 3-variables correlation, as we did above (see Appendix \ref{appInt}). The larger is $N$, the more accurate the approximation is. However, for the present scope, the quality provided by the substitution~(\ref{split}) is sufficient.

Within this approximation, letting
$\Delta = 4\left \{\left [1+(2\tau)^{-1}\right ]^2-u\left (\mu\tau^{-1}+u\right )\right \}$,
Eq.~(\ref{eqE}) has, for $\tau$ sufficiently large as to have $\Delta >0$, the following solution
\be
E(t)= E(\infty)-\frac{\sqrt{\Delta}}{2u}\left[1-\tanh \left (\frac {\sqrt \Delta }{2}\,t+k\right )\right],
\ee
where:
\be
k=\tanh^{-1}\left(\frac{2uE(0)}{\sqrt \Delta}+\frac{2}{\sqrt{\Delta}}\left(1+\frac{1}{2\tau}\right)\right) , \quad 
E(\infty)=-\frac{1}{u}\left (1+\frac{1}{2\tau}\right )+\frac{1}{2u}\sqrt \Delta .
\ee
This form approaches exponentially the asymptotic value $E(\infty)$ after a characteristic time 
\be
t_{stat}=\frac {2}{\sqrt \Delta }\simeq \left [1-u^2+\frac{1}{2}(1-u\mu)\tau^{-1}\right ]^{-1},
\label{tstat}
\ee
the last passage holding for large $\tau$. 
This expression shows that
a new timescale
\be
\tau_c\simeq \frac{1-u\mu}{2(1-u^2)}
\label{tauc}
\ee
exists, separating two regimes where one or the other term on the r.h.s. of Eq.~(\ref{tstat}) prevail, such that $t_{stat}\simeq (1-u^2)^{-1}$ for $\tau \gg \tau_c$
or $t_{stat}\simeq \frac{2\tau}{1-u\mu}$ for $\tau \ll \tau_c$.
In the former case, stationarization 
occurs on a time $t_{stat}$ which is independent of $\tau$ and is due to the thermal flipping of spins $\{S_i\}$. 
In this case $t_{stat}$ amounts to the relaxation time of the corresponding quenched model.
In the latter case, $t_{stat}$ is proportional to $\tau$ and time-translational invariance is induced
by the flipping of the $\{\sigma_i\}$.
Notice that this is always the case at $B=\infty$, since $u=1$, where the corresponding quenched model ages. 
The above scenario, with two different
stationarization mechanisms separated by $\tau_c$, is perfectly analogous
to the one discussed before for model i),
but the novelty here is that $\tau_c$ can be tuned large at will by increasing $B$.

Let us now study the behavior of two-time quantities in the stationary state.
Using the generalization of the fluctuation-dissipation theorem in~\cite{Lippiello_2005,Corberi_2007,PhysRevB.77.212201,PhysRevE.78.041120,Baiesi_2009,PhysRevE.81.011124} to non-equilibrium states, similarly to the paramagnet case, the response function can be cast as 
\be
B^{-1}R(s)= -\frac{dC(s)}{ds}+
\frac{u}{4}[D(s)+\widetilde D(s)],
\label{Rinteract}
\ee
where
\be
D=\langle S_{i+1} (s) \Delta \sigma_i (s)S_i(0) \rangle,
\quad \quad
\widetilde D(s)=\langle S_i (s) \Delta \sigma_i(s)S_{i+1} (0)\rangle
\ee
are usually called {\it asymmetry} terms
and $\Delta\sigma_i(s)=\sigma_i(s)-\sigma_i(0)$. However, similarly to $E(t)$, the equations for $C,D,\widetilde D$ are not closed. Despite this, supplementing Eq.~(\ref{Rinteract}) with some physical arguing, it is possible to discuss the thermalization properties.

To start, it is clear that the behavior of $C$ and $R$ must be different if $\tau\gg\tau_c$ or $\tau\ll\tau_c$.
In the former case, since decorrelation is due to the thermal flipping of the $\{S_i\}$, we expect
$C(s)\equiv C_q^{(eq)}(s)$, where $C_q^{(eq)}(s)$ is the equilibrium form of $C(s)$ in the quenched model~\cite{Godreche} which is, clearly, independent of $\tau$. Similarly, we also expect $R(s)\equiv R_q^{(eq)}(s)=-B\frac{dC_q^{(eq)}(s)}{ds}$, $R_q^{(eq)}(s)$ being the equilibrium response of the quenched model, which obeys the equilibrium fluctuation dissipation theorem with respect to $C^{(eq)}_q$. The same result is arrived at upon inspection of Eq.~(\ref{Rinteract}) because there is no variation $\Delta \sigma_i$ on the timescale $t_S$ where $C(s)$ drops to zero and hence the asymmetry terms in Eq.~(\ref{Rinteract}) vanish.\\
The above considerations hold for finite $s$. Letting $s\to \infty$ it is clear that also the $\sigma_i$ at some point will start to evolve, making $D$ and $\widetilde D$ finite. On such huge timescale we expect thermalization with the slow bath to occur. Hence we argue the following form for $X$
\begin{equation}
    X(C)=\left \{
    \begin{array}{lll}
    \frac{\beta}{B}\quad , &\mbox{for} & C\equiv 0 \\
    1\quad , &\mbox{for} &0<C\le 1.
\end{array}
\right .
\label{fdtplot_interact1}
\end{equation}
This result can be checked by numerical simulations where $C,D,\widetilde D$ are
measured directly and, from them, $R$ is obtained through Eq.~(\ref{Rinteract}).
The result is shown in Fig.~\ref{fig_fdtplot_ferromagnet1}, where we plot $B^{-1}\chi$ against $C$, similarly to what done before for model i). 
On the left panel one sees that, increasing $\tau$, the curves tend to
the limiting behavior~(\ref{fdtplot_interact1}), valid for $\tau\gg \tau_c$, for any finite value of $C$ (for the values of $B$ and $\beta$ considered, from Eq.~(\ref{tauc}) one has $\tau_c\simeq 1.19$, in this case). Regarding the isolated point $C=0$, this is clearly not accessible in simulations and, therefore, the first line of Eq.~(\ref{fdtplot_interact1}) remains a reasonable conjecture.

In the right panel of Fig.~\ref{fig_fdtplot_ferromagnet1} one observes the effect of lowering the temperature of the $\{S_i\}$. 
While for sufficiently high temperature
(e.g. for $B=0.1$) one has $\tau\gg \tau_c$ and the form~(\ref{fdtplot_interact1}) holds,
as $B$ is increased the condition
$\tau \gg \tau_c$ is no longer met
(for $B=1$, for instance, it is $\tau_c\simeq 7.08$),
and a different, non trivial fluctuation-dissipation
relation, to be discussed soon, is observed.

\begin{figure}[htbp]
	\centering
		\includegraphics[width=0.49\textwidth]{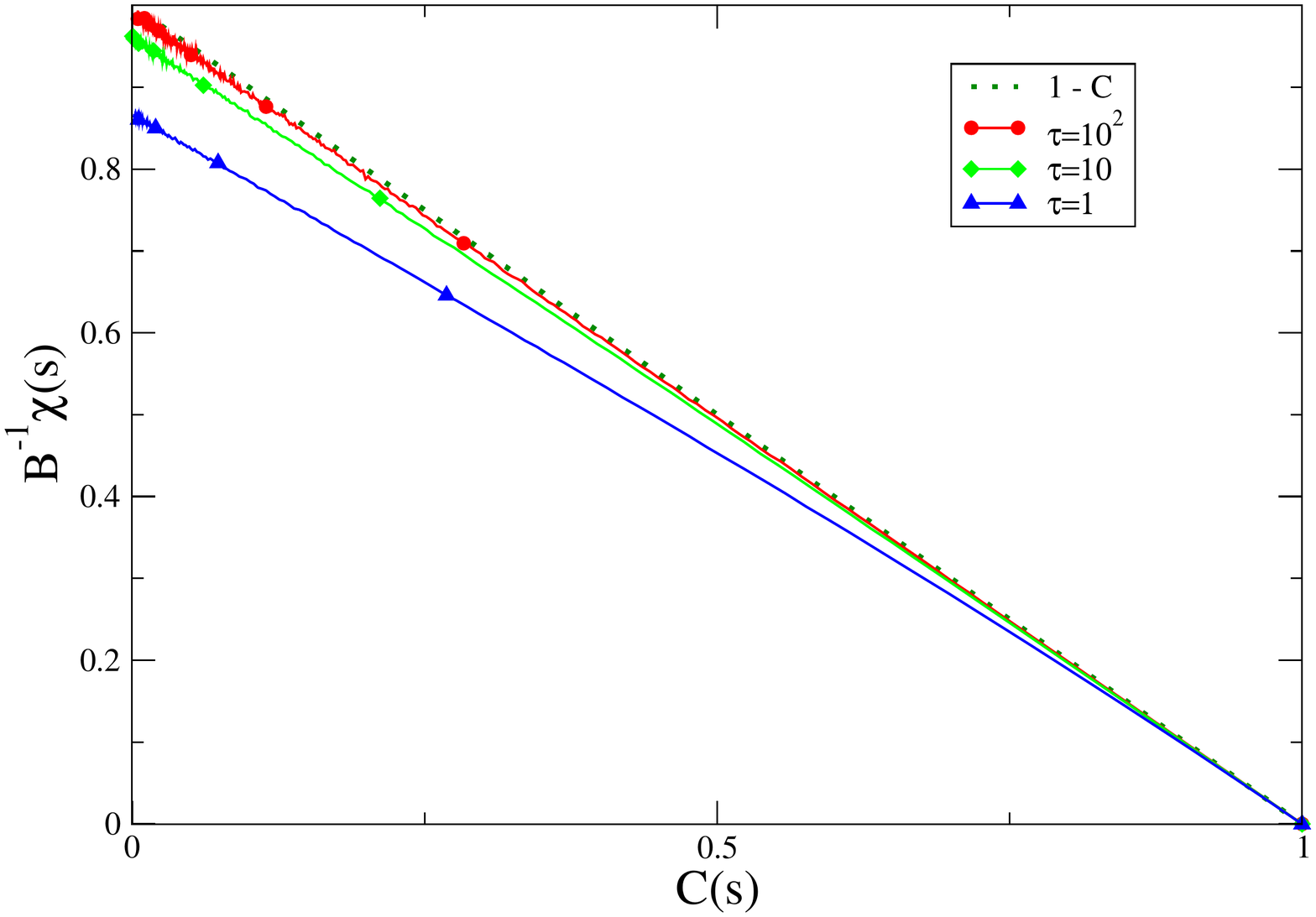}
        \includegraphics[width=0.49\textwidth]{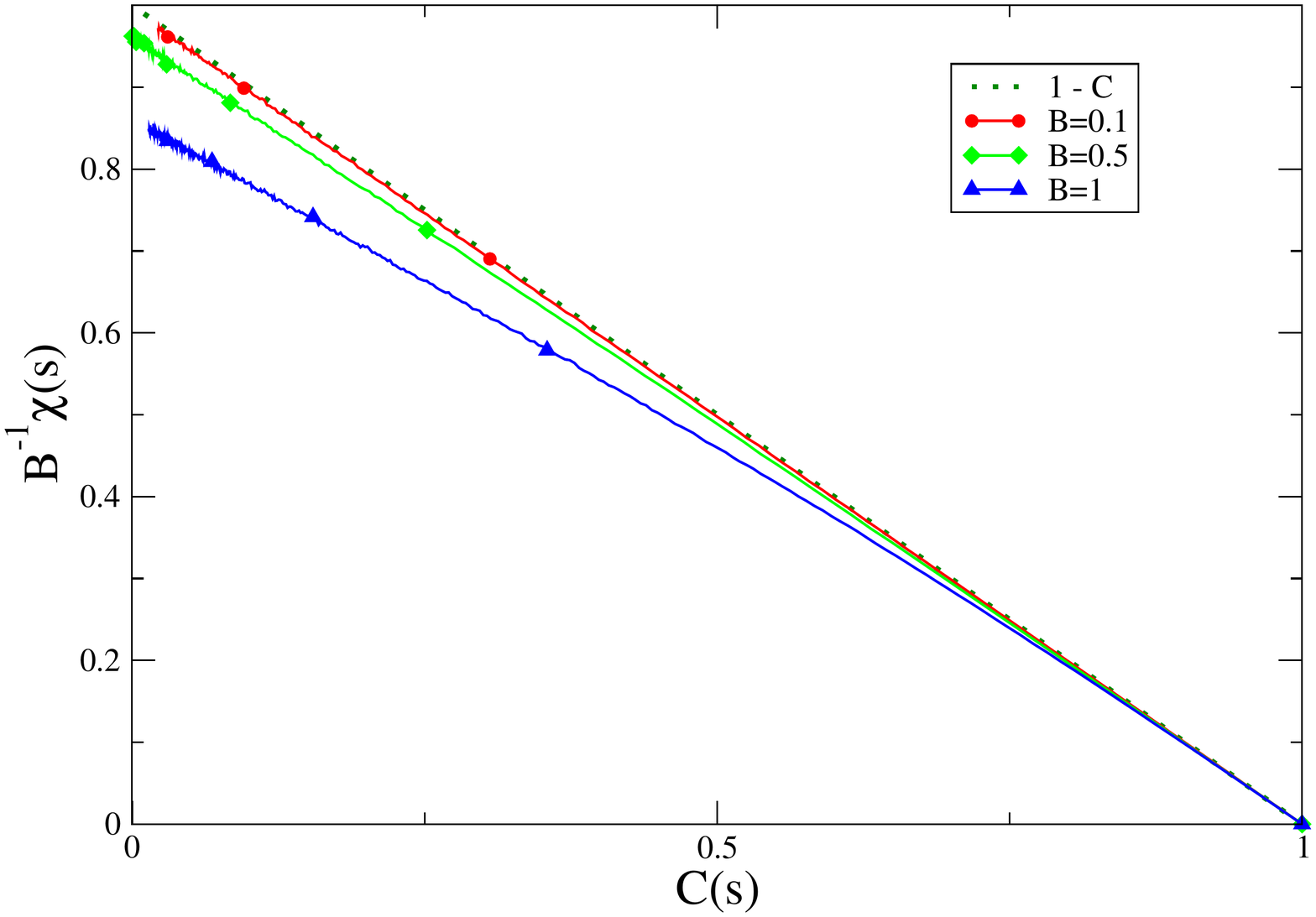}
	\caption{The transition towards the regime described by equation~(\ref{fdtplot_interact1}).  On the left, $B^{-1}\chi(s)$ is plotted against $C(s)$ for $B=0.5$, $\beta=0$ and different values of $\tau$. On the right,   the same plot is shown for $\tau=10$, $\beta=0$ and different values of $B$.}
	\label{fig_fdtplot_ferromagnet1}
\end{figure}

Next we consider the case in which $\tau\ll\tau_c$. At variance with the previous situation, now decorrelation of the $\{S_i\}$ is brought about by the flipping 
of the $\{\sigma _i\}$, hence we expect $C$ to decay on a timescale of order $t_{\sigma S}$ which is larger the larger is $\tau$, so the asymmetry terms $D,\widetilde D$ can never be neglected in Eq.~(\ref{Rinteract}).
In order to understand what happens in this case one has to investigate the physical mechanisms whereby the stationary state decorrelates. The nature of such state can be more easily inferred at $B=\infty$, $\beta=0$.
In this case, in the quenched model, all the local interactions are satisfied, i.e. $E\equiv -1$. However,
flipping of the $\{\sigma _i\}$ in the annealed system causes some local interaction to be unsatisfied. We will say that there is a kink, or an interface, in the position where this occurs. 
Interfaces can diffuse without energy cost and, upon meeting, they annihilate. The equality between the kink production and annihilation rates is the mechanism whereby their number is kept constant at stationarity. 

The larger is $\tau$, the smaller is the fraction of kinks created, hence the longer is the time needed for their mutual annihilation.
Then, on timescales much smaller than $\tau$, in the stationary state, interfaces are neither produced nor annihilated, they can only diffuse. This is exactly what happens in the corresponding quenched model~\cite{Corberi_2019} when it is aging long after a temperature quench, provided that, also in this case, we restrict to times where kink annihilation does not occur. Therefore we expect $C(s)= C_q^{(ag)}(s,t_k)$ and 
$R(s)=R_q^{(ag)}(s,t_k)$,
where $C_q^{(ag)}(s,t_k)$ is the correlation function of the corresponding quenched model, aging after a temperature quench, computed at a time $t_k$ when the density of interfaces is the same as in the stationary state of the annealed model, and similarly for $R_q^{(ag)}(s,t_k)$. Explicit expressions for $C_q^{(ag)}$ and $R_q^{(ag)}$ have been derived in~\cite{Lippiello-Zannetti,Godreche}, which correspond to the form of $X(C)$ in the second line of Eq.~(\ref{fdtLipZan}).
According to that, the plot of $\chi$ vs $C$ is the same as that found in the aging system quenched to $B=\infty$.
This is shown in the inset of the right panel of Fig.~\ref{fig_fdtplot_ferromagnet2}.
\begin{figure}[htbp]
	\centering
		\includegraphics[width=0.49\textwidth]{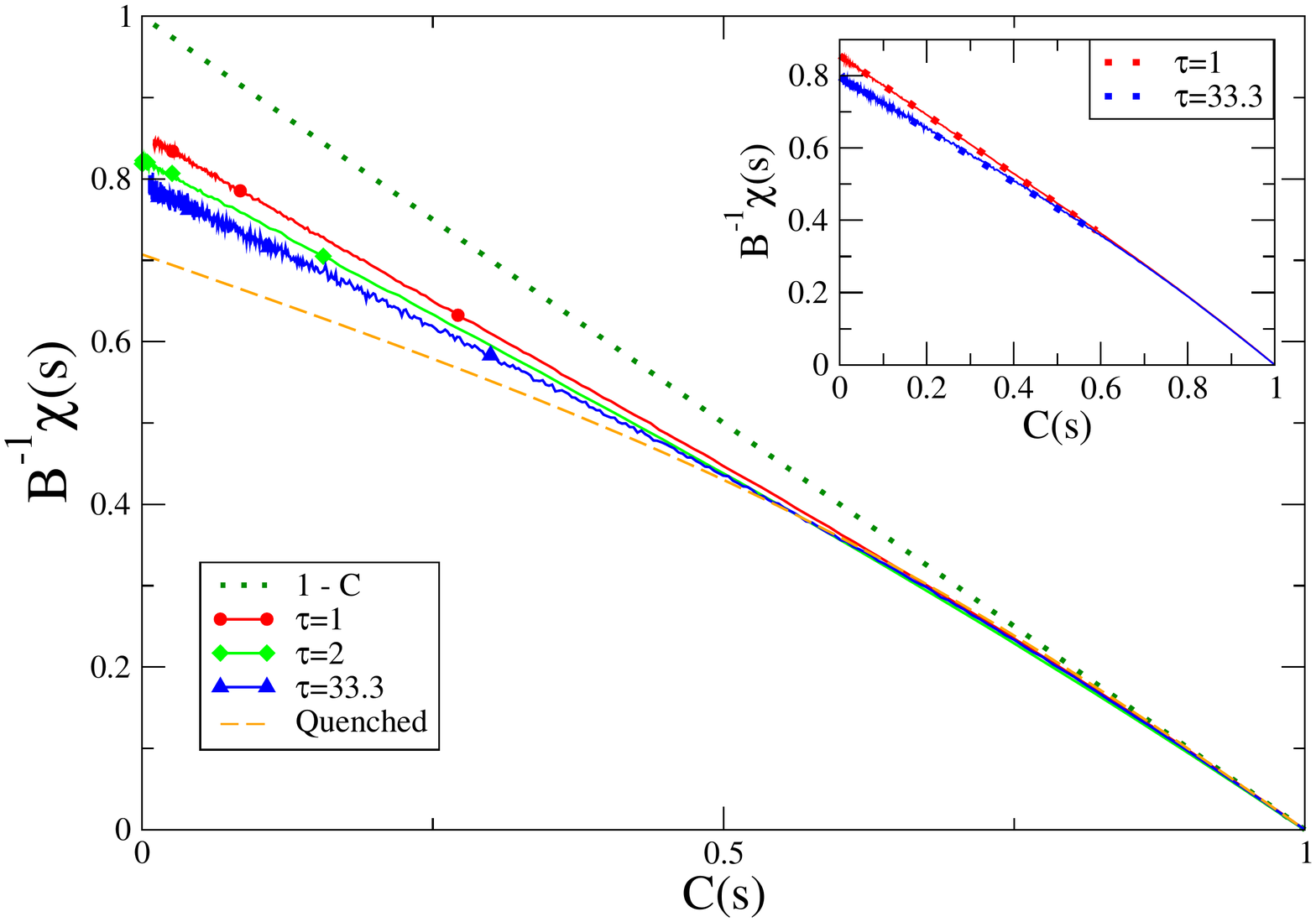}
		\includegraphics[width=0.49\textwidth]{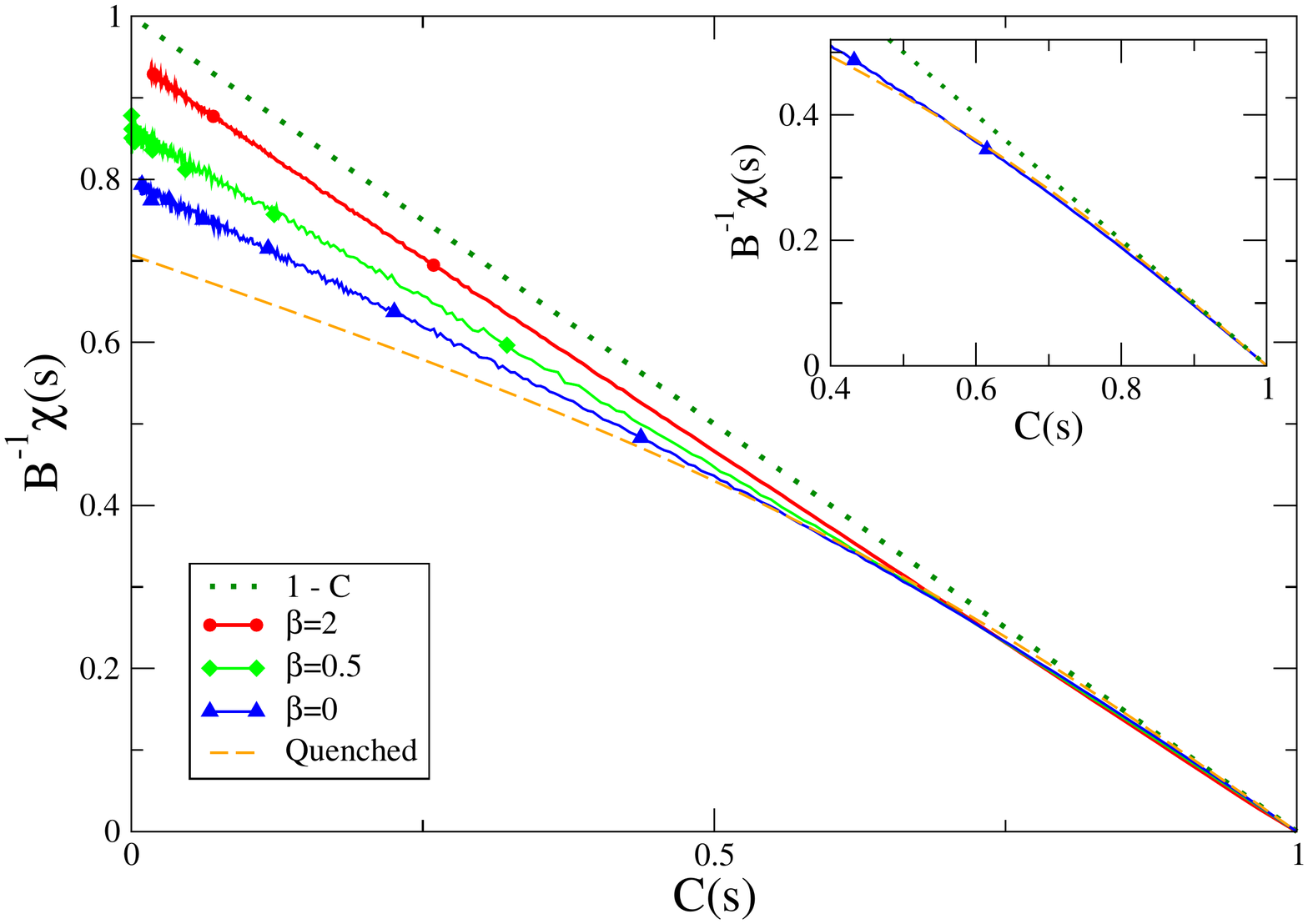}
 \caption{$B^{-1}\chi(s)$ is plotted against $C(s)$ for $B=\infty$, $\beta=0$ and different values of $\tau$ (left panel) or for $\tau=20$ and different $\beta$ (right panel). Increasing $\tau$ beyond $\tau=20$ does not change the shape of the curve.
 The inset of the left panel shows a comparison of the numerical data with the lines $B^{-1}\chi(C)=1-X(C^*)C$ (heavy-dotted lines, see main text) for two values of $\tau$ ($\tau=1$ and $\tau=33.3$). 
 The inset in the right panel is a zoom of the case $\beta=0$.}
	\label{fig_fdtplot_ferromagnet2}
\end{figure}

As said, in the time domain
$s\ll \tau$ considered insofar,
no new kinks are created, neither annihilations occur, $C$ and $R$ vary as due only to their movement. When $s$ becomes of order $\tau$, instead, corresponding to a certain value $C^*\simeq C(s\simeq\tau)$ of the correlation, interfaces start to be both created and annihilated. Here the equivalence with the quenched model in the aging state is broken, because in the latter kinks can only be annihilated. Since we have no intuition on the mechanism whereby the response is built in this regime, we have to fully resort, in this case, to numerical simulations.
From this study, whose main results are reported in Fig.~\ref{fig_fdtplot_ferromagnet2} (left panel), evidence emerges that $X(C)$ remains
approximately constant and equal to the value $X^*=X(C^*)$
in the whole time range with $s>\tau$
(namely the curve $\chi (C)$ in Fig.~\ref{fig_fdtplot_ferromagnet2} is a straight line from $C=C^*$ down to $C=0$). A detailed numerical check of this is contained in the inset of the left panel of Fig.~\ref{fig_fdtplot_ferromagnet2}.
In order to obtain this figure we have
evaluated the 
analytic form in the second line of Eq.~(\ref{fdtLipZan}) in the point $C^*$ where we observe the numerical curve to depart from it and we have plotted the line $B^{-1}\chi(C)=1-X^*C$. The agreement of this line with the numerical data is very good. Notice that this seems to hold for any value of $\tau$ (see the two lines drawn in figure).
Hence we find the following form
\begin{equation}
    X(C)=\left \{
    \begin{array}{lll}
    \left [2-\sin^2\left (\frac{\pi}{2}C^*\right )\right ]^{-1}\quad , &\mbox{for} &0\le C< C^* \\
\left [2-\sin^2\left (\frac{\pi}{2}C\right )\right ]^{-1}\quad , &\mbox{for} & C^*\le C\le 1,
\end{array}
\right .
\label{fdtLipZan}
\end{equation}
Notice that, despite there is a well defined value of $X=X^*$ in the whole range of $0\le C\le C^*$, this value has nothing to do with the temperature of the slow bath.
However, the existence of a unique value $X^*$ of the fluctuation-dissipation ratio in a whole range
of $C$ is remarkable and calls for a physical interpretation that presently is missing.
Conversely, the temperature of the fast bath can be read, but only at $C=0$, similarly to what observed for the paramagnetic model with $B=\infty$ (see Sec.~\ref{paramagnet},
Eq.~(\ref{X_para_T0})), for similar reasons. 
Hence in this case multi-thermalization is not observed. The modification of the above scenario for finite $\beta$ are shown in the right panel of Fig.~\ref{fig_fdtplot_ferromagnet2}, still with $B=\infty$.
Increasing $\beta$ it is
clear that the system approaches the equilibrium condition 
at $B=\beta=\infty$. Hence we expect the 
fluctuation-dissipation plot to approach the equilibrium 
form~(\ref{equiR}), as it is actually observed in the figure. 

\section{Conclusions} \label{conclusions}

In this paper we have studied the issue of systems in contact with a multibath, in the context of two simple paradigmatic magnetic models where explicit calculations can be fully or partly carried out. In such models, two sets of Boolean variables $\{S_i\}$ and $\{\sigma_i\}$, fast and slow respectively, are attached to two thermal baths at different temperatures. The dual models obtained by quenching the slow variables $\{\sigma_i\}$ represent, respectively, a paramagnetic
system where the $\{S_i\}$ do not interact and an interacting spin-glass model. In this study we focused mainly on the properties of the stationary state. Our interest was on the thermalisation properties, namely how energy transfer between the system and the baths occurs on different timescales, which is well encoded into the fluctuation-dissipation relation between the correlation and the response function of the fast variables $\{S_i\}$. From this, the effective (inverse) temperature ${\cal B}=BX(s)$,
where $X(s)$ is the fluctuation-dissipation ratio on timescale $s$ and $B^{-1}$ is the temperature
of the bath coupled to the fast variables, can be inferred.

Our analysis showed that, while for the simpler {\it non-interacting} model i) the thermalization 
properties agree with what one would expect on the basis of a thermodynamically inspired intuition,
the situation is more involved in model ii).
Specifically, in the regime $\tau\gg 1$, when the $\{\sigma _i\}$ become extremely slow, in model i) timescales are sharply separated into two sectors by $\tau$: for $s\lesssim \tau$, 
heat exchanges between the $\{S_i\}$ and the multibath are regulated by the temperature $B^{-1}$
of the {\it fast bath}, whereas for $s\gtrsim \tau$ the relevant temperature is the one of the {\it slow bath} $\beta^{-1}$. Instead, in model ii), for sufficiently low temperature $B^{-1}$ one finds a fully non-trivial result. On short timescales $s\lesssim \tau$, $X(s)$ exhibits a continuous dependence on $s$, 
which is the same as the one observed in the aging one-dimensional Ising model quenched to zero 
temperature~\cite{Lippiello-Zannetti,Godreche,CLZ2001, PhysRevE.65.046114}. Let us stress again, however, that what we find
in the annealed model pertains to its stationary state. On larger timescales $s\gtrsim \tau$ there is a well defined fluctuation-dissipation ratio $X^*$ whose simple interpretation in physical terms is now missing. In this respect it is worth recalling
that the above-mentioned non-trivial $X(s)$ found in the aging Ising model has been shown~\cite{Sollich_2002} not to possess the correct thermodynamical properties to be associated to a {\it true} effective temperature, specifically it is not observable-independent.  This suggests that one should be cautious with simple thermodynamically-inspired descriptions of 
multibath systems, particularly when interactions among degrees of freedom may play an important role.

The studies conducted in this paper leave a number of open questions behind, which are possible
matter for future investigations.
Perhaps the most relevant one regards the generality of our results. In particular, one wonders 
if other models might share some of the properties of the ones studied here. In this respect, let us observe, in passing, that the shape of the fluctuation-dissipation plot of model ii) closely resembles the one observed in multibath mean-field spin-glasses~\cite{Contucci_2021}. Another relevant question regards the study of the fluctuation relation \cite{Lebowitz_1999,Jarzynski_2004} in such types of models: while in model i) we expect a trivial result, in model ii) the  temperature associated to heat exchanges with the thermal baths could lead to a highly non trivial phenomenology possibly related to our result for the fluctuation-dissipation ratio.

\appendix
\section{Evolution equation for correlators}
The starting point of all the computations are the following results \cite{Glauber}. The equal-time average of Boolean variables evolve according to
\be
\frac{d}{d t}\langle \xi_{i_1}(t)\ldots  \xi_{i_m}(t) \rangle=-2\langle \xi_{i_1}(t)\ldots  \xi_{i_m}(t) \sum^m_{k=1}w(\xi_{i_k}(t)) \rangle \ ,
\label{startingpointeq}
\ee
where $\xi_i$ can be either a variable of kind $\{S_i\}$ or $\{\sigma_i\}$ and $w(\xi_{i_k}(t))$ is the transition rate of a flip $\xi_i(t)\rightarrow -\xi_i(t)$, which in our case is given by Eqs. (\ref{rate1}-\ref{rate2}). The generalization to a two-time correlator is
\be
\frac{d}{d s}\langle \xi_{i_1}(t+s)\ldots  \xi_{i_m}(t+s) \xi_{j_1}(t)\ldots \xi_{j_n}(t) \rangle=-2\langle \xi_{i_1}(t+s)\ldots  \xi_{i_m}(t+s) \xi_{j_1}(t)\ldots \xi_{j_n}(t) \sum^m_{k=1}w(\xi_{i_k}(t+s)) \rangle \ .
\label{startingpoint}
\ee

\section{Paramagnet}
\label{apppara}
\subsection{Evolution equations and fluctuation-dissipation plot}
The Glauber rates in the case of the paramagnet are given by (where $m=\tanh B$ and $\mu=\tanh \beta$)
\be
\omega _{S_i}(\{S _i\},\{\sigma_i\})=\frac{1}{2}\left (1-m \sigma_i S_i \right ) \ , \quad \quad
\omega _{\sigma_i}(\{S _i\},\{\sigma_i\})=\frac{1}{2\tau}\left (1-\mu \sigma_i S_i \right ) \ .
\label{paramrates}
\ee
Specializing Eq.~(\ref{startingpointeq}) to the case of the energy $E(t)=-\langle S_i(t) \sigma_i(t) \rangle$, using rates~(\ref{paramrates}), equation~(\ref{evoenergy}) of the main text is obtained.\\

Using Eq.~(\ref{startingpoint}), the evolution equation for the two-time functions are given by
\be
\begin{aligned}
\frac{d C(t,s) }{ds}&=-C(t,s) +m A(t,s)  \\
\frac{d A(t,s)}{ds}&=-\frac{1}{\tau}A(t,s) +\frac{\mu}{\tau}C(t,s)
\end{aligned} \quad \quad
\begin{aligned}
\frac{d\tilde A(t,s)}{dt}&=-\tilde A(t,s) +m\Sigma(t,s)  \\
\frac{d\Sigma(t,s)}{dt}&=-\frac{1}{\tau}\Sigma(t,s) +\frac{\mu}{\tau}\tilde A(t,s) \ ,
\end{aligned}
\label{sist2time}
\ee
where $C(t,s)$, $A(t,s)$, $\tilde A(t,s)$ and $\Sigma(t,s)$ have been defined in Eq. (\ref{defC}) and after Eqs. (\ref{sistC}), (\ref{resp}) and (\ref{sistA}), respectively.
The evolution equations in Eqs.~(\ref{sist2time}) coincide with Eqs.~(\ref{sistC},\ref{sistA}). The proper initial conditions are $C(t,0)=\Sigma(t,0)=1$, while the initial condition on $A(t,0)=\tilde A(t,0)$ must be obtained from the solution of the evolution equation of $-\langle S_i(t) \sigma_i(t) \rangle$, which is analogous to Eq.~(\ref{evoenergy}), up to a minus sign in the constant term on the right hand side. Solving this system one has
\be
\begin{aligned}
&C(t,s)=\beta_{C1}(t) e^{-\frac{s}{t_S}}+\beta_{C2}(t) e^{-\frac{s}{t_{\sigma S}}}\\
&A(t,s)=\beta_{A1}(t) e^{-\frac{s}{t_S}}+\beta_{A2}(t) e^{-\frac{s}{t_{\sigma S}}}\\
&\tilde A(t,s)=\beta_{\tilde A1}(t) e^{-\frac{s}{t_S}}+\beta_{\tilde A2}(t) e^{-\frac{s}{t_{\sigma S}}} \ ,
\end{aligned}
\label{twopointexplicit}
\ee
where $t_S$ and $t_{\sigma S}$ are defined in the main text (below Eq.~(\ref{Cstat})) and the quantities $\beta_a$ are reported below, in sub-section~\ref{appexpl}.
From the first of Eqs.~(\ref{twopointexplicit}), taking the limits $\tau\rightarrow\infty$, $t\rightarrow\infty$, one gets Eq.~(\ref{Cstat}).\\

The response can be found from the general result in~\cite{Lippiello_2005}, adapted to our notation,
\be
B^{-1}R(t,s)=\frac{1}{2}\left(\frac{dC(t,s)}{dt}-2\frac{dC(t,s)}{ds}-m\widetilde A(t,s)-mA(t,s)\right ).
\label{respersp}
\ee
In the stationary case the correlation is independent of $t$ and this reduces to Eq. (\ref{resp}). The integrated response is found in the stationary case from Eq. (\ref{chi}). Fixing the temperatures and $\tau$, the first of Eqs. (\ref{twopointexplicit}) with $B^{-1}\chi(s)$ give a parametric curve in the fluctuation-dissipation plot, varying $s$, which are exactly the curves of Fig. \ref{fig_fdtplot_ferromagnet1}.

\subsection{Explicit expressions of the coefficients appearing in Eq.~(\ref{twopointexplicit})} 
\label{appexpl}
The various $\beta$'s in Eq.~(\ref{twopointexplicit}) have these expressions (to simplify the notation we have set $\alpha\equiv\sqrt{(\tau-1)^2+4\tau m \mu}$):
\be
\begin{aligned}
\beta_{C1}(t)=\frac{e^{-\frac{t(1+\alpha+\tau)}{\tau}}}{2 \alpha^{2}(1+\tau)}\left(e^{t+\frac{t}{\tau}}(-1+\alpha+\tau)\left(-1+\tau^{2}\right)+2 \tau\left(-\alpha+e^{t+\frac{t}{\tau}}(2-\alpha+2 \tau)\right) \mu m-2\left(-1+e^{t+\frac{t}{\tau}}\right) \alpha \tau^{2} m^{2}\right)
\end{aligned}
\ee
\be
\begin{aligned}
\beta_{C2}(t)=\frac{e^{-\frac{t(1+\tau)}{\tau}}}{2 \alpha^{2}(1+\tau)}\left(e^{t+\frac{t}{\tau}}(-1-\alpha+\tau)\left(-1+\tau^{2}\right)+2 \tau\left(-\alpha+e^{t+\frac{t}{\tau}}(2+\alpha+2 \tau)\right) \mu m+2\left(-1+e^{t+\frac{t}{\tau}}\right) \alpha \tau^{2} m^{2}\right)
\end{aligned}
\ee
\be
\begin{aligned}
\beta_{A1}(t)=&\frac{1}{2 \alpha^{2}(1+\tau)} e^{-\frac{t(1+\alpha+\tau)}{ \tau}}\left(e^{\frac{t \alpha}{\tau}}\left(-1+e^{t+\frac{t}{\tau}}\right)(-1+\tau) \tau(-1-\alpha+\tau) m+4 e^{\frac{t \alpha}{\tau}}\left(-1+e^{t+\frac{t}{\tau}}\right) \tau \mu^{2} m+\right. \\
&\left.\mu\left(e^{\frac{t \alpha}{\tau}}(1+\alpha-\tau)(-1+\tau)+e^{\frac{t(1+\alpha+\tau)}{\tau}}\left((-1+\tau)^{2}-\alpha(1+3 \tau)\right)+4 e^{\frac{t \alpha}{\tau}}\left(-1+e^{t+\frac{t}{\tau}}\right) \tau^{2} m^{2}\right)\right)
\end{aligned}
\ee
\be
\begin{aligned}
\beta_{A2}(t)=&\frac{1}{2 \alpha^{2}(1+\tau)} e^{-\frac{t(1+\tau)}{ \tau}}\left(\left(-1+e^{t+\frac{t}{\tau}}\right)(-1+\tau) \tau(-1+\alpha+\tau) m+4\left(-1+e^{t+\frac{t}{\tau}}\right) \tau \mu^{2} m+\right. \\
&\left.\mu\left(-(-1+\tau)(-1+\alpha+\tau)+e^{t+\frac{t}{\tau}}\left(\alpha+(-1+\tau)^{2}+3 \alpha \tau\right)+4\left(-1+e^{t+\frac{t}{\tau}}\right) \tau^{2} m^{2}\right)\right)
\end{aligned}
\ee
\be
\begin{aligned}
\beta_{\tilde A1}(t)=&\frac{1}{2 \alpha^{2}(1+\tau)} e^{-\frac{t(1+\alpha+\tau)}{\tau}}\left(\tau\left(-e^{\frac{t \alpha}{\tau}}(-1+\tau)(-1+\alpha+\tau)+e^{\frac{t}{(1+\alpha+\tau)}}\left((-1+\tau)^{2}-\alpha(3+\tau)\right)\right) m+\right. \\
&\left.4 e^{\frac{t}{\tau}}\left(-1+e^{t+\frac{t}{\tau}}\right) \tau \mu^{2} m+e^{\frac{t \alpha}{\tau}}\left(-1+e^{t+\frac{t}{\tau}}\right) \mu\left((-1+\tau)(-1+\alpha+\tau)+4 \tau^{2} m^{2}\right)\right)
\end{aligned}
\ee
\be
\begin{aligned}
\beta_{\tilde A2}(t)=
&\frac{1}{2 \alpha^{2}(1+\tau)} e^{-\frac{t(1+\tau)}{ \tau}}\left(\tau\left((1+\alpha-\tau)(-1+\tau)+e^{t+\frac{t}{\tau}}\left((-1+\tau)^{2}+\alpha(3+\tau)\right)\right) m+\right. \\
&\left.4\left(-1+e^{t+\frac{t}{\tau}}\right) \tau \mu^{2} m+\left(-1+e^{t+\frac{t}{\tau}}\right) \mu\left(-(1+\alpha-\tau)(-1+\tau)+4 \tau^{2} m^{2}\right)\right) \ .
\end{aligned}
\ee

\section{Spin glass}
\label{appInt}
\subsection{Evolution equations}
Glauber rates for this model are (where $u=\tanh (2B)$ and $\mu=\tanh \beta$)
\be
\omega _{S_i}(\{S _i\},\{\sigma_i\})=\frac{1}{2}\left[1-\frac{u}{2}S_i(\sigma_i S_{i+1}+\sigma_{i-1}S_{i-1})\right]  \ , \quad \quad
\omega _{\sigma_i}(\{S _i\},\{\sigma_i\})= \frac{1}{2\tau}\left(1-\mu \sigma_iS_iS_{i+1}\right) \ .
\label{paramrates1}
\ee
Using these expressions in Eqs.~(\ref{startingpoint}), we cannot obtain closed evolution equations for any set  of observable quantities of some interest.
For example, let us consider the equal-time correlators
\be
G_k=\langle S_i \sigma_i \sigma_{i+1} \cdots \sigma_{i+k-1} S_{i+k}\rangle
\label{evoG}
\ee
(with the convention $G_0=1$) and the ones
\be
H^{i,l,\ldots}_{k,m,\ldots}=\langle S_i \sigma_i \sigma_{i+1}\cdots  \sigma_{i+k-1} S_{i+k}S_l \sigma_l \sigma_{l+1}\cdots  \sigma_{l+m-1} S_{l+m} \cdots \rangle \quad \text{for} \quad k,m\ge1 \ , \quad l>i+k \ , \quad ...
\label{generalcorrelator}
\ee
obtained from them by repeating an arbitrary number of times the sequence of variables appearing in Eqs.~(\ref{evoG}) (here all the numbers $k,m,\ldots$ are positive natural while $i,l,\ldots$ are integer). Notice that all these correlators are invariant under the gauge transformation~\cite{Nishimori}
\be
S_i\longrightarrow S_i \psi_i \, \quad \quad \sigma_i\longrightarrow \sigma_i \psi_i \psi_{i+1}
\ee
where $\psi_i=\pm 1$, therefore, we call the correlators of type  (\ref{generalcorrelator}) (and, as a consequence, also the ones of type (\ref{evoG})) {\it correlators involving gauge variables}.\\

Using Eq.~(\ref{startingpoint}) in the case $G_k$ ($k>0$) and the translational invariance property
\be
\langle \xi_{i_1+j}\cdots \xi_{i_n+j}\rangle=\langle \xi_{i_1}\cdots \xi_{i_n}\rangle \ ,\quad \forall j \in \mathbb Z \ ,
\label{transinv}
\ee
with $\xi_i$ generic Boolean variable, one gets the evolution equation (for $k\ge 1$)
\be
\begin{aligned}
\frac{dG_k}{dt}=&-2\left( \frac{k}{2\tau}+1\right)G_k+\left(\frac{2-\delta_{k,1}}{\tau}\mu+u\right)G_{k-1}+u G_{k+1}+\frac{\mu}{\tau}\sum_i Q^k_i
\end{aligned}
\label{evogauge}
\ee
where $\delta_{a,b}$ is the Kronecker delta and $Q^k_i$ are all the possible correlators involving gauge variables of type $H^{i,{i+m+1}}_{m,n}$ with $m+n=k-1$, according to the definition (\ref{generalcorrelator}).
For example the cases for the lowest possible $k$'s are:
\begin{itemize}
\item for $k=1,2$, no correlator of type $Q^{k}$ exists since the constraints $n+m=0$ or $n+m=1$ do not admit solutions for $m\ge 1$ and $n\ge 1$;
\item for $k=3$ the constraint give $m=n=1$ so the only possible correlator is $Q^{k=3}_1=\langle S_i\sigma_{i} S_{i+1} S_{i+2}\sigma_{i+2} S_{i+3}\rangle$;
\item for $k=4$ the constraint give $m=1$ and $n=2$ or vice versa so we have the cases
\be
Q^{k=4}_1=\langle S_i\sigma_{i}S_{i+1}S_{i+2}\sigma_{i+2}\sigma_{i+3}S_{i+4}\rangle \quad \quad Q^{k=4}_2=\langle S_i\sigma_{i}\sigma_{i+1}S_{i+2}S_{i+3}\sigma_{i+3}S_{i+4}\rangle .
\ee
\end{itemize}
Notice that, by comparing Eq.~(\ref{eqE}) and Eq.~(\ref{evoG}), $G_1(t)=-E(t)$. This way, setting $k=1$ in Eq.~(\ref{evogauge}) with the convention $G_0=1$, one gets exactly Eq.~(\ref{eqE}).\\

\subsection{Correlators tending to zero at long times and stationarization}
As already discussed for the model at hand, Eqs.~(\ref{startingpoint}) are an infinite set of coupled linear equations, with the correlators as unknown variables. 
Despite this infinite hierarchy, however, it is still possible to argue which quantities attain a finite value at stationarity.  In order to do that, let us notice that the only dynamical equation for the correlators presenting a constant term is the one for $G_1$. To see this, let us consider the l.h.s of Eq.~(\ref{startingpointeq}) and notice that it has the structure of a product of the variables with the corresponding rates. As a consequence, the only manner to generate a constant term is to use the Boolean property of the variable to substitute an equal-time product of two variables with a 1. Now, notice that in Glauber rates (\ref{paramrates1}) the Boolean variables appear only in the combination $S_i\sigma_i S_{i+1}$ (up to translations, for Eq.~(\ref{transinv})). This way, substituting the rates in Eq.~(\ref{startingpointeq})), the only manner to have a constant term is that the correlator contains only $S_i\sigma_i S_{i+1}$, because this generates a constant in the dynamical equation thanks to the Boolean property. But this is exactly $G_1$.\\
Secondly, repeating the computation made to obtain Eq.~(\ref{evogauge}) in order to generalize it to general correlators involving gauge variables $H^{i,l,\ldots}_{k,m,\ldots}$, it is easy to see that in this case all the evolution equations contain \textit{only} correlators involving gauge variables.
This last property is crucial to understand which correlators vanishes for long times, since the linear system of differential equation is split into two blocks, one involving correlators $H^{i,l,\ldots}_{k,m,\ldots}$ with a constant term and another involving all the others without any constant term. This implies that all the correlators but $H^{i,l,\ldots}_{k,m,\ldots}$ tend to zero at long times. The stationarization value of the correlators can be found by putting to 0 all the derivatives with respect to time, resulting in a block system of linear equations: one block containing only correlators involving gauge variables (and as a consequence also all the $G_k$) with a constant term, with solution which is different from zero and another block containing all the other correlators, with null solution. To conclude, all the correlators are null at stationarization, but the correlators involving gauge variables, $H^{i,l,\ldots}_{k,m,\ldots}$.

\subsection{The splitting approximation}
We briefly discuss here the splitting approximation introduced in Eq.~(\ref{split}). Since in the limit $\tau\rightarrow \infty$ the rate for the $\sigma$'s vanishes (see Eq.~(\ref{paramrates1})), one recovers the original Glauber dynamics, where only the $S$'s evolve. In this case, it has been shown~\cite{Glauber} that $\langle S_i S_{i+k}\rangle=|\eta|^k$, with $\eta$ solving the equation $\eta^2+2u^{-1}\eta+1=0$. This implies that $\langle S_i S_{i+k}\rangle=\langle S_i S_{i+1}\rangle^k$. Multiplying this expression by $\sigma^k$, which is constant in this case, and inserting a set of squares of Boolean variable one gets
\be
\langle S_i \sigma S_{i+1} S_{i+1} \cdots S_{i+k-1} S_{i+k-1} \sigma S_{i+k}\rangle=\langle S_i \sigma S_{i+1}\rangle^k \ ,
\ee
which is exactly the splitting approximation Eq.~(\ref{split}). This also implies that in Glauber model, that is the limit $\tau\rightarrow\infty$, this approximation becomes exact.
Similarly, the approximation becomes exact also in the limit $\beta\rightarrow \infty$, $B\rightarrow \infty$. This is because with $m=u=1$ the constant solution of Eq.~(\ref{evogauge}) is $G_n=Q_n=1$ for all $n$'s, which trivially implies the splitting:
\be
G_n=G_{n-l}G_l
\ee
for all $n$ and $l$, such that $1\le0<m$. For $m=2$ and $l=1$ one gets Eq.~(\ref{split}).

\bibliography{bibliografia}

\end{document}